\documentclass[pra,twocolumn,graphicx]{revtex4}
\usepackage{amssymb}
\usepackage{epsfig,amsmath}

\begin{document}

\title{Controlling high-harmonic generation and above-threshold ionization with an
attosecond-pulse train}
\author{C. Figueira de Morisson Faria$^{1}$, P. Sali\`eres$^2$, P. Villain$^2$,
and M. Lewenstein$^{3,4}$} \affiliation{$^1$Centre for Mathematical
Science, City University,
Northampton Square, London EC1V 0HB, United Kingdom\\
$^2$CEA-SPAM, B\^{a}t. 522, Centre d'Etudes de Saclay, F-91191
Gif-Sur-Yvette, France\\
$^3$ICREA, Instituci\'{o} Catalana
di Recerca i Estudis Avan\c{c}ats, and ICFO, Institut de Ci\`encies Fot\`oniques, E-08860 Castelldefels (Barcelona), Spain\\
$^4$Institut f\"ur theoretische Physik, Universit\"at Hannover,
Appelstr. 2, D-30167 Hannover, Germany}

\date{\today}

\begin{abstract}
We perform a detailed analysis of how high-order harmonic generation
(HHG) and above-threshold ionization (ATI) can be controlled by a
time-delayed attosecond-pulse train superposed to a strong,
near-infrared laser field. In particular we show that the
high-harmonic and photoelectron intensities, the high-harmonic
plateau structure and cutoff energies, and the ATI angular
distributions can be manipulated by changing this delay. This is a
direct consequence of the fact that the attosecond pulse train can
be employed as a tool for constraining the instant an electronic
wave packet is ejected in the continuum. A change in such initial
conditions strongly affects its subsequent motion in the laser
field, and thus HHG and ATI. In our studies, we employ the
Strong-Field Approximation and explain the features observed in
terms of interference effects between various electron quantum
orbits. Our results are in agreement with recent experimental
findings and theoretical studies employing purely numerical methods.
\end{abstract}
\maketitle
\section{Introduction}
 High-frequency light pulses of
attosecond ($10^{-18}s$) duration have caused a breakthrough in
metrology, allowing one to resolve and control dynamical processes
occurring in the subfemtosecond and subangstr\"om scale
\cite{AdiM2004}. This is in particular possible due to the fact that
one attosecond is roughly the time it takes for light to travel
through atomic distances. Attosecond pulses may be used, for
instance, for tracing the motion of bound electrons \cite{atto1},
exciting inner shell electrons \cite{atto2}, controlling molecular
motion \cite{atto3}, electron emission \cite{atto4}, high-harmonic
generation \cite{atthhg2004,dpg} or above-threshold ionization
\cite{attati2005}.

Such pulses owe their existence to the fact that high-order
harmonics, which are generated by rare gases interacting with intense $(I\sim
10^{14}\mathrm{W/cm^2})$, near-infrared (IR) laser fields of
femtosecond duration, are nearly phase locked. Hence, by superposing a
large number of harmonics one may obtain attosecond pulses in the
extreme ultraviolet range \cite{AdiM2004}. There exist two approaches to
attosecond-pulse production. One may generate isolated
attosecond pulses from few-cycle laser pulses
\cite{attoexp,fewcyclerev,atto1}, or attosecond-pulse trains
from laser pulses comprising several cycles \cite{attoexptrain}.
Such attosecond pulse trains have been predicted in the late 1990's
\cite{saclay96}, and experimentally realized a few years later
\cite{attoexptrain}. Since then, there has been considerable
improvement in their contrast, duration and peak intensity.
In particular, an intrinsic chirp of the attosecond emission
was evidenced, setting a lower limit on the duration of the attosecond
pulses obtained by grouping an increasing number of harmonic orders \cite{mairesse}.
Thereby, important issues are an adequate choice of the group of harmonics,
and optimal propagation conditions in the gaseous media. In this
context, it is of interest to control the atto-chirp and more generally
the intensity-and order-dependent phase of the atomic dipole \cite{optimatto}.

Recently, it has been shown that an attosecond-pulse train
superposed to an intense laser field can be used for manipulating
high-order harmonic generation (HHG) \cite{atthhg2004,dpg} and
above-threshold ionization (ATI) \cite{attati2005}. In particular,
the resolution, intensity and maximal energies of both ATI and HHG
spectra turned out to be strongly dependent on the time delay
between the laser field and the attosecond pulse train. This
behavior has been predicted theoretically
\cite{atthhg2004,attati2005} and observed experimentally in both
cases \cite{attati2005,dpg}. It has been interpreted in the light of
the physical mechanisms governing both phenomena.

In HHG, an electron reaches the continuum by tunneling or
multiphoton ionization at a time $t^{\prime}$, propagates in the
continuum and, subsequently, at a time $t$, recombines with its
parent ion \cite{tstep}. In this process, the kinetic energy
acquired by this electron is released in form of high-frequency
radiation. Similarly, for ATI, an electron leaves an atom, either
reaching the detector, or elastically recolliding with it. This
leads to the low- or high-order ATI peaks, respectively. The
attosecond pulse train can be used for controlling the ejection of
such an electron in the continuum. Specifically, the time
$t^{\prime}$ can be fixed by changing the time delay between the
attosecond pulse train and the IR field. This constraint upon the
ionization time can be used as a tool for selecting a particular
orbit of an electron reaching the continuum \cite{atthhg2004,dpg},
or a particular momentum transfer from the IR field to the electron
\cite{attati2005}. As a direct consequence, the ATI and HHG spectra
can be manipulated.

In this paper, we investigate the effects reported in
\cite{atthhg2004,dpg,attati2005} for HHG and ATI in detail, using a
quantum-mechanical formulation of the above-stated processes
\cite{hhgsfa,atisfa,orbitshhg}, within the Strong-Field Approximation (SFA)
\cite{footnsfa}. This approach has some advantages over that
employed in \cite{atthhg2004,attati2005}, i.e., the numerical
solution of the time-dependent Schr\"{o}dinger equation (TDSE).
Indeed, even though the SFA involves a higher degree of modeling and
physical approximations, it is much less demanding, from the
computational viewpoint. This is particularly important for a
realistic description of the macroscopic response of the gaseous
system, in which the single-atom response serves as input for
propagation codes. Furthermore, it provides a transparent and clear
physical interpretation, in terms of the orbits of an electron
leaving or recombining with its parent ion. This is a further
advantage over the TDSE, for which it may be difficult to extract
the physical mechanisms involved.

In this work, we put a particularly strong emphasis on the quantum
interference between the different possible paths of an electron
ejected by the attosecond-pulse train. Specifically, we investigate
how the initial conditions with which the electron reaches the
continuum influence such an interference, and which consequences
this has on the ATI or HHG spectra. We show that the attosecond
pulses can be used for controlling this interference, and explain
the features observed in \cite{atthhg2004,dpg,attati2005} in terms
of electron orbits.

This paper is organized as follows: In Sec.~\ref{theo}, we briefly
discuss the SFA for ATI and HHG, with emphasis on the changes
introduced by the attosecond-pulse train. Subsequently (Sec.
\ref{res}), we show how several features of both phenomena are
affected by it, and provide physical explanations. Finally, in Sec.
\ref{concl}, we give a summary of such results and state the main
conclusions of this work.

\section{Transition amplitudes}
\label{theo}
\subsection{Above-threshold ionization}
 For ATI, we will concentrate on the direct
electrons, which reach the detector without rescattering with their
parent ion. In this case, the ATI transition amplitude reads
\begin{equation}
b_p(t)=i\int_{0}^{t}\mathbf{E}(t^{\prime })dt^{\prime }d_{z}(\mathbf{p}+\mathbf{A}%
(t^{\prime }))e^{iS(\mathbf{p},t,t^{\prime })},\label{atiampl}
\end{equation}with the action
\begin{equation}
S(p,t,t^{\prime })=-\int_{t^{\prime }}^{t}\left( \frac{\left[
p+A(t^{\prime \prime })\right] }{2}^{2}+I_{p}\right) dt^{\prime
\prime }.\label{action1}
\end{equation}

Eq. (\ref{atiampl}) can be understood as follows: At an instant
$t^{\prime}$,  an electron is ejected by the external laser field
from the atomic state in which it is initially bound, reaching a
continuum state. It then remains in the continuum until a time $t$,
when it reaches the detector with final momentum $\mathbf{p}$. In
Eq. (\ref{atiampl}), $\mathbf{A}(t)$, $I_p$, and
$\mathbf{E}(t')=-d\mathbf{A}(t')/dt'$ denote the vector potential, the atomic
ionization potential, and the electric field at the time of its
release, respectively. We take the dipole matrix element
$d_z(\mathbf{p}+\mathbf{A}(t))=\left\langle \mathbf{p}+\mathbf{A}(t)\right| \mathbf{r}%
.\mathbf{\epsilon }_{z}\left|\psi _{0} \right\rangle$ for an initial
$1s$ state, where the vector $\mathbf{\epsilon }_{z}$ denotes the
polarization axis. Specifically, this yields
\begin{equation}
d_z(\mathbf{p})=i \left(\frac{2^{19/4}I_p^{5/4}}{\pi}\right)
\frac{\mathbf{p}.\mathbf{\epsilon
}_{z}}{[\mathbf{p}^2+2I_p]^3}
\label{dip1s}
\end{equation}

 We will now write the transition amplitudes for the specific case in
 which an attosecond-pulse train is present.
 We assume that this train is composed of an infinite number of odd
harmonics. This yields
\begin{equation}
\mathbf{E}_{\mathrm{atto}}(t)=E_{h}\pi \sum_{n=0}^{\infty
}\frac{(-1)^{n}}{\sigma(t)}\delta
(t-\frac{n\pi}{\omega})\mathbf{\epsilon}_{z}, \label{attotrain}
\end{equation}
where $\omega$, $E_{h}$, and $\sigma(t)$ denote the
laser field frequency, the attosecond-pulse strength and the train temporal
envelope, respectively. In this paper, we assume that the envelope is
either given by
\begin{equation}
\sigma(t)\sim \exp[\zeta |t|], \label{attowidth}
\end{equation}
where $\zeta\ll 1$, or by $\sigma=const.$ Physically, the latter
case corresponds to an infinitely long attosecond-pulse train. We
approximate the laser field by a monochromatic wave
$\mathbf{E}_l(t)=-d\mathbf{A}_l(t)/dt$, with the vector potential
$\mathbf{A}_l(t)=2\sqrt{U_p}\cos(\omega t-\phi)\mathbf{\epsilon
}_{z}$ of frequency $\omega$ dephased of $\phi$ with respect to the
attosecond-pulse train.

We now assume that the attosecond pulse train releases the electron
in the continuum and that its subsequent propagation is determined
only by the monochromatic field. This implies that
$\mathbf{E}(t')\simeq \mathbf{E}_{\mathrm{atto}}(t')$ and
$\mathbf{A}(t)\simeq \mathbf{A}_l(t)$ in Eq.~(\ref{atiampl}). For
the direct ATI electrons, such assumptions lead to the transition
amplitude
\begin{eqnarray}
b_p&\sim&[d_{+}\cos \alpha -id_{-}\sin \alpha] \chi (a^{(o)},\zeta
\pi/\omega)\nonumber\\&&+[d_{-}\cos \alpha -id_{+}\sin \alpha] \chi
(a^{(e)},\zeta \pi/\omega), \label{atifinite}
\end{eqnarray}
where $ d_{\pm }=d_z(\mathbf{p}+\mathbf{A}(t^{\prime }))\pm
d_z(\mathbf{p}-\mathbf{A}(t^{\prime }))$, $\theta$ gives the angle
between the laser field and the final electron momentum, and the
argument $\alpha =2p\sqrt{U_{p}}/\omega \cos \theta \sin \phi$ comes
from the non-trivial part of the action. Furthermore, in Eq.
(\ref{atifinite}),
\begin{equation}
\chi (a,\zeta \pi/\omega)=\frac{\sinh(\zeta
\pi/\omega)}{-\cos(a)+\cosh(\zeta \pi/\omega)}, \label{peakedfct}
\end{equation}
$a^{(o)}=(I_p+U_p+p^2/2-\omega)\pi/\omega$ and
$a^{(e)}=(I_p+U_p+p^2/2)\pi/\omega$, respectively. Eq.
 (\ref{peakedfct}) exhibits sharp peaks at $\cos a=1$ (i.e., for
$a=2n\pi$). Specifically, $a^{(o)}$ and $a^{(e)}$ lead to
expressions with steep maxima around the odd and even ATI peaks,
respectively.

 In the limit of infinitely long attosecond-pulse trains,
 $b_p=b^{(e)}_p+b^{(o)}_p$, where
\begin{equation}
b^{(e)}_{p}\sim\sum_{n}\left[ d_{-}\cos \alpha -id_{+}\sin \alpha
\right] \delta (\frac{p^{2}}{2}+U_{p}+I_{p}-2n\omega
)\label{atieven}
\end{equation}
and
\begin{equation}
b^{(o)}_{p}\sim \sum_{n}\left[ d_{+}\cos \alpha -id_{-}\sin \alpha
\right] \delta (\frac{p^{2}}{2}+U_{p}+I_{p}-(2n+1)\omega
),\label{atiodd}
\end{equation}
gives the contributions from the even and odd ATI peaks,
respectively \cite{footnfinite}. In Eq.~(\ref{atieven}) and
(\ref{atiodd}), one can explicitly notice that the electron leaves
with vanishing momentum if $2n\omega=(I_p+U_p)$. Physically, this
means that the ionization potential $I_p$ is effectively shifted by
the ponderomotive potential $U_p$. Such an effect is noticeable for
relatively high driving-field intensities.

In some of the subsequent results, in order to perform a more direct
comparison with the results existing in the literature, we will
integrate the photoelectron yield over the solid angle. In this
case, the yield will be given by
\begin{equation}
\eta (p)=2 \pi \int^1_{-1}|b_{p}|^2 d(\cos \theta), \label{angleint}
\end{equation}
where $b_{p}$ is given by Eq.~(\ref{atifinite}).

Furthermore, apart from initial 1s states, we will consider the
so-called broad gaussian (GBR) approximation, for which
$d_z(\mathbf{p}+\mathbf{A}(t))=const$. Physically, this
approximation means that the electron is initially in an extremely
localized bound state (for details see, e.g., \cite{hhgsfa}). In
this particular case, and in the limit of infinitely long
attosecond-pulse trains, the integral (\ref{angleint}) can be solved
analytically and gives
\begin{equation}
\eta (p)\sim 1\pm \frac{\sin(2\alpha_0)}{2\alpha_0},
\label{angleintGBR}
\end{equation}
where $\alpha_0=2p\sqrt{U_{p}}/\omega\sin \phi$, and the positive
and negative signs correspond to the odd and even ATI peaks,
respectively.

\subsection{High-harmonic generation}

The HHG transition amplitude is given by an expression which is
slightly different from Eq. (\ref{atiampl}), namely
\begin{eqnarray}
b_{\Omega}&\hspace{-0.1cm}=\hspace*{-0.1cm}&i\int_{-\infty }^{\infty
}\hspace*{-0.5cm}dt\int_{-\infty }^{t}~\hspace*{-0.5cm}dt^{\prime
}\int d^{3}kd_z^{\ast
}(\mathbf{k}+\mathbf{A}(t))d_z(\mathbf{k}+\mathbf{A}(t^{\prime
}))\nonumber \\&& E(t^{\prime })\exp [iS(t,t^{\prime },\Omega
,\mathbf{k})], \label{amplhhg}
\end{eqnarray}
with the action
\begin{equation}
S(t,t^{\prime },\Omega ,\mathbf{k})=-\frac{1}{2}\int_{t^{\prime }}^{t}[%
\mathbf{k}+\mathbf{A}(\tau )]^{2}d\tau -I_{p}(t-t^{\prime })+\Omega
t. \label{actionhhg}
\end{equation}
Eq. \ref{amplhhg} describes a physical process in which an electron
is freed at a time $t^{\prime}$, propagates in the continuum with
momentum $\mathbf{k}$ from $t^{\prime}$ to $t$, and, at this time,
recombines with its parent ion, generating a harmonic of frequency
$\Omega$.

We will now apply the same physical assumptions as in the previous
sections to the transition amplitude (\ref{amplhhg}). We will,
however, adopt a slightly different convention: instead of
considering the time delay in the infra-red field, we will take
$\mathbf{A}_l(t)=2\sqrt{U_p}\cos(\omega t)e_z$ and insert the delay
$t_d=\phi/\omega$ in the attosecond pulses. Such a convention will facilitate
the discussions in Sec.~III.B, in terms of electron orbits. Under
such assumptions, Eq. (\ref{amplhhg}) reads
\begin{eqnarray}
b_{\Omega}&=&\frac{i\pi E_{\mathrm{h}}}{\sigma}\sum_{n=0}^{\infty}
(-1)^{n}\int_{-\infty }^{+\infty
}\hspace*{-0.5cm}dt\int d^{3}k\exp \left[ iS(t,t^{\prime}_n,\Omega,\mathbf{k})\right] \nonumber \\
&& d_z^{\ast }(%
\mathbf{k}+\mathbf{A}_l(t))
 d_z(\mathbf{k}+\mathbf{A}_l(t^{\prime}_n)).
 \label{hhgatto}
\end{eqnarray}
In this case, the ionization time is being fixed at $t^{\prime
}_n=t_{d}+n\pi /\omega $, with $n$ an integer, because the electron
is being released into the continuum at such an instant by the
attosecond pulses.

The transition amplitude is further simplified in the sense that
only the integrals in $\mathbf{k}$ and $t$ must be solved. We
evaluate this equation employing saddle-point methods. For that
purpose, we consider $\mathbf{k}$ and $t$ so that
$S(t,t^{\prime},\Omega,\mathbf{k})$ is stationary, i.e., that
$\partial S/\partial \mathbf{k}=\mathbf{0}$ and $\partial S/\partial
t=0$. This yields the saddle-point equations
\begin{equation}
\int_{t^{\prime }_n}^{t}d\tau \left[ \mathbf{k}+\mathbf{A}_l(\tau
)\right] =0, \label{saddle3}
\end{equation}
and
\begin{equation}
2(\Omega -I_{p})=\left[ \mathbf{k}+\mathbf{A}_l(t)\right] ^{2},
\label{saddle2}
\end{equation} whose solutions can be directly
associated to the classical orbits of an electron returning to [Eq.
(\ref{saddle3})] and recombining with [Eq. (\ref{saddle2})] its
parent ion \cite{orbitshhg}. In the presence of the attosecond-pulse
train the saddle-point equations for $t$ and $\mathbf{k}$ can be
combined, so that the electron return time is given by
\begin{eqnarray}
&&\sin \omega t-(-1)^{n}\sin \omega t_{d}\\
&=&\left[ \omega (t-t_{d})-n\pi)\right] \left(\cos \omega t\mp
\sqrt{(\Omega -I_{p})/(2U_p)}\right). \nonumber \label{saddleatto}
\end{eqnarray}
 For details on the approximation employed we refer to Ref.
\cite{atiuni}.

Alternatively, one may follow the procedure in Ref.~\cite{hhgsfa},
and compute the Fast Fourier transform for the time-dependent dipole
moment in the Strong-Field Approximation. Such an approach involves
a single saddle point approximation in the intermediate electron
momentum, and two numerical integrations. Hence, it possesses the
advantage of taking into account all sets of orbits, and not only
the above-mentioned pairs. The procedure adopted in this paper,
however, allows a more detailed assessment of quantum-interference
effects. For comparison, we provide the time-dependent equations
employed in \cite{hhgsfa}, together with a brief discussion, in the
appendix.

\section{Results}

\label{res}
\subsection{Above-threshold ionization}
 As a starting point, we will investigate the effect reported in
\cite{attati2005}, in which the yield extends to much higher
energies for vanishing delay. In \cite{attati2005}, this is
justified by the following argument: if the electron leaves the atom
at the zero of the IR field (i.e., at $\phi=n\pi$), its vector
potential is at is maximum, i.e., $A(t')=\pm 2 \sqrt{U_p}$. Hence,
when the electron reaches the continuum, under these conditions it
acquires maximum possible drift momentum. Therefore, it reaches the
detector with higher energies. Since this effect only influences how
the electron is ejected in the continuum, this should affect both
direct and rescattered electrons equally. Therefore, we expect that
Eq.~(\ref{atieven}) and (\ref{atiodd}), for the direct electrons,
exhibit such a behavior.

\begin{figure}[tbp]
\begin{center}
\includegraphics[width=8cm]{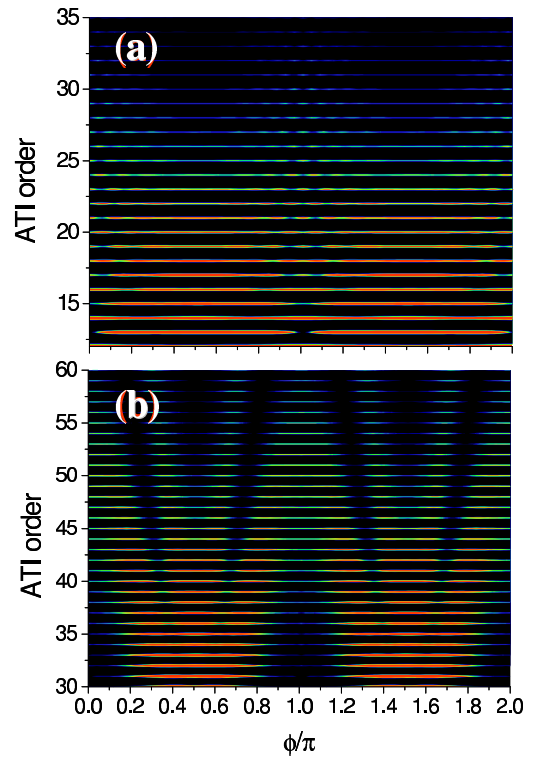}
\end{center}
\caption{ (Color online) Direct above-threshold ionization spectra,
integrated over the solid angle and in linear scale (Eq.
(\ref{angleint})), for argon ($I_p=0.58$ a.u.) interacting with a
monochromatic field of frequency $\omega=0.057$ a.u. and an
attosecond-pulse train (Eq. (\ref{attotrain})), as functions of the
time delay between both driving fields.
 Panels (a) and (b) have been computed for an intensity of
$I=3.3\times 10^{13} \mathrm{W/cm^2}$ and $I=5\times 10^{14}
\mathrm{W/cm^2}$ for the low-frequency laser field, respectively. In
panels (a) and (b), the ATI orders (labelled with the number of
photons from the ground state) correspond to photoelectron energies
roughly between 0 and 36.5 eV, and 0 and 47.4 eV, respectively, due
to the ponderomotive shift. The yield has been computed from Eq.
(\ref{atifinite}), with $\zeta=0.04$. The intensities of the
attosecond pulse trains have been chosen as $I_h=I/10$, i.e., one
tenth of those of the low-frequency field. } \label{atifig1}
\end{figure}

 Fig.~\ref{atifig1} displays ATI spectra
obtained from Eq.~(\ref{atifinite}), for time delays $0 \leq\phi\leq
2\pi$. In order to perform a direct comparison with the results in
\cite{attati2005}, we integrate the yield over the solid angle [Eq.
(\ref{angleint})]. In the upper part of the figure
[Fig.~\ref{atifig1}.(a)], we consider the same parameters as in
\cite{attati2005}, while in the lower part [Fig.~\ref{atifig1}.(b)]
 we choose a higher intensity for the IR field \cite{footnote}. In
the figure, there exist two distinct behaviors, within two energy
regions. For photoelectron energies near the low-energy end of the
spectra, the peaks exhibit maximal intensities near
$\phi=(2n+1)\pi/2$, i.e., if the electron is released close to the
times for which the electric field is maximal. For higher energies,
the yield presents a clear minimum around such a delay, and the
spectra extends to much higher energy ranges if the delay is
vanishing, i.e., if the electron reaches the continuum at the
electric-field crossings. This behavior agrees with the findings in
\cite{attati2005}, and becomes more extreme as the laser-field
intensity increases.

Both energy regions appear to be dependent on the driving-field
intensity. For instance, for Fig.~\ref{atifig1}.(a), maxima near
$\phi=(2n+1)\pi/2$ are observed roughly for ATI orders $12\lesssim
N\lesssim 20$ ($0\lesssim p^2/2\lesssim 13.3eV$), while for
Fig.~\ref{atifig1}.(b), this region lies within $32\lesssim
N\lesssim 40$ ($0\lesssim p^2/2\lesssim 16.4eV$). Beyond such energy
regions, the yield exhibit minima for such delays. Apart from that,
for high-energy photoelectrons, there exist additional minima for
the higher driving-field intensity [c.f. Fig.~\ref{atifig1}.(b)].
%%%%%%%%%%%%%%%%%%%%%%%%%%%%%%%%%%%%%%%%%%%%%%%
%%%%%%%%%%%%%%%%%%%%%%%
%%%%%%%%%%%%%%%%%%%%%%%%%%%%%%%
\begin{figure}[tbp]
%\begin{center}
\hspace*{-1cm}
\includegraphics[width=9.5cm]{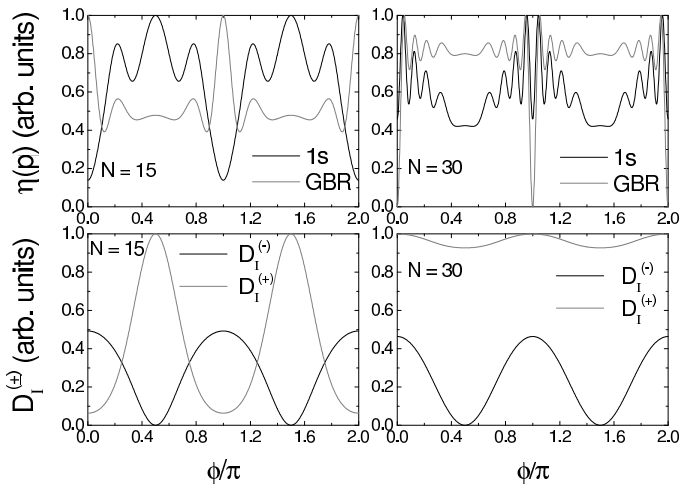}
%\end{center}
\caption{Upper panels: modulus squared of the transition amplitudes
(\ref{atieven}) and (\ref{atiodd}), integrated over the solid angle,
for 1s initial states (black line) and within the broad gaussian
approximation (grey line), as functions of the time delay between
the attosecond-pulse train and the IR field. Lower panels:
$|d_{+}|^2$ and $|d_{-}|^2$ integrated over the solid angle, for
$1s$ initial states. We considered argon under the influence of an
IR field of $I=3.3\times 10^{13}\mathrm{W/cm^2}$ superposed to an
attosecond-pulse train of $I_h=I/10$ (Panel (a) in Fig. 2). The
curves in the upper panels have been normalized so that their
maximum values are unity, and those in the lower panel to the
maximum value of $D^{(-)}_I$. The numbers N in the figure denote the
ATI orders.} \label{atifig2}
\end{figure}
%%%%%%%%%%%%%%%%%%%%%%%

Such a behavior can be readily seen from Eqs.~(\ref{atieven}) and
(\ref{atiodd}), or, more specifically, from the angle-integrated
yield [Eq.(\ref{angleint})]. Indeed, we expect $\eta_p$ to be the
superposition of two types of contributions, coming respectively
from the non-trivial part of the action $\alpha$ and from the dipole
matrix elements $d\pm$. Physically, the former corresponds to the
phase the electron acquired leaving at a particular instant, and the
latter are specific to the initial state in which the electron is
bound.

 In Fig.~\ref{atifig2}, we
display the isolated contributions from such terms, integrated over
the solid angle [Eq. (\ref{angleint})], for distinct ATI orders. In
order to disentangle the effect due to the dipole matrix element and
that due to the non-trivial part of the action, in addition to
considering 1s initial states, we computed spectra in the broad
gaussian (GBR) approximation, for which
$d_z(\mathbf{p}+\mathbf{A}(t))=const$ [Eq. (\ref{angleintGBR})]. In
this case, $d_{+}=1$ and $d_{-}=0$, so that the effect caused by the
non-trivial part of the action, i.e., $\sin \alpha$ and $\cos
\alpha$ in Eq.~(6), can be directly visualized (c.f. gray curves in
the upper panels).

 We observe that the non-trivial part of the
action leads to a highly oscillating pattern, which, for even or odd
ATI peaks, have maxima or minima at $\phi=n\pi$, respectively. As
the electron energy increases, the oscillating region becomes
increasingly flatter, and the amplitude of the oscillations are
always larger near $\phi=n\pi$. It is, however, clear that the
action, by itself, is not responsible for the behavior exhibited in
Fig.~\ref{atifig1} and shown in detail in Fig.~\ref{atifig2}, for
initial $1s$ states (c.f. black curves in the upper panels). In
fact, even though this term leads to several extrema, it does not
cause an overall enhancement or suppression of the yield near
$\phi=n\pi$, as observed for the low- and high-order ATI peaks
respectively.

The integral
\begin{equation}
D^{(\pm)}_I=2 \pi\int^{1}_{-1}|d_{\pm}|^2d(cos\theta)
\end{equation}
of $|d_{\pm}|^2$ over the solid angle sheds some light in this
behavior (c.f. lower panels in Fig.~\ref{atifig2}). The
contributions from $d_{-}$ exhibit a minimum at $\phi=(2n+1)\pi/2$,
while the behavior of those from $d_{+}$ depend on the energy region
in question. Near the ionization threshold, the latter contributions
exhibit maxima near the delays $\phi=(2n+1)\pi/2$. As the ATI order
increases, such maxima become flatter, until the concavity of
$D^{(+)}_I$ changes and the yield starts to exhibit maxima at
$\phi=n\pi$. Such contributions are dominant, with respect to those
from $d_{-}$, and are present both in Eq.~(\ref{atieven}) and
(\ref{atiodd}). This behavior gets more pronounced as the
driving-field intensity increases (higher intensities lead also to
sets of secondary maxima in $D^{(\pm)}_I$, which contribute to the
additional structure observed in Fig.~\ref{atifig1}.(b)). One should
also note that, very near the threshold, for $\phi=n\pi$,
$D^{(-)}_I>D^{(+)}_I$. This causes the minima observed for $\eta(p)$
in Fig.~1(a), for initial $1s$ states. In contrast, the yield from
the broad-gaussian approximation exhibits maxima in this case.

Therefore, the behavior reported in \cite{attati2005} is not only
due to the phase the electron obtains when it is ejected in the
continuum by the attosecond-pulse train, or, in other words, to the
momentum transfer from the field to the electron. Additionally, it
is related to the initial state in which the electron is bound,
which, in our model, is a $1s$ state. Indeed, such a behavior is
absent in the broad gaussian approximation, which, physically,
corresponds to an electron initially bound in a highly localized
state. Such a strong dependence on the initial electronic bound
state is, in a sense, quite unusual for phenomena occurring in the
context of atoms in strong laser fields.

%%%%%%%%%%%%%%%%%%%%%%%%%%%%%%%
\begin{figure}[tbp]
\hspace*{-1cm}
\includegraphics[width=9.5cm]{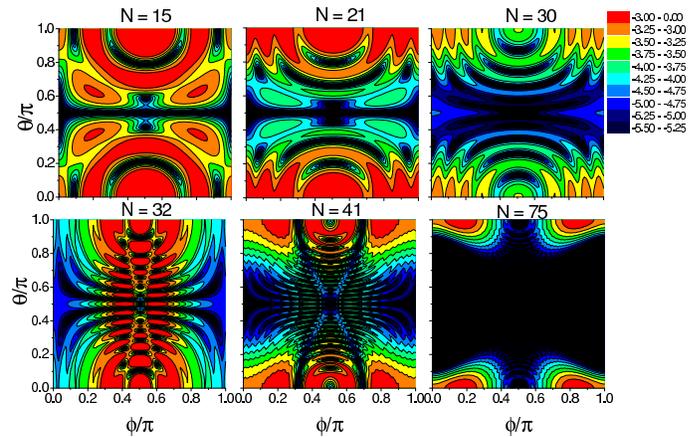} \caption{ (Color
online) Angular distributions for the ATI direct electrons, in a
logarithmic scale, for the same field and atomic parameters as in
Fig.~1. The upper and lower panels have been computed for an
intensity of $I=3.3\times 10^{13} \mathrm{W/cm^2}$ and $I=5\times
10^{14} \mathrm{W/cm^2}$ for the low-frequency laser field,
respectively. The yields are displayed as functions of the time
delay $\phi=\omega t_d$ and orientation angle $\theta$ between the
laser field and the final electron momenta. The ATI orders N are
indicated on the upper parts of the panels, and all contours have
been normalized to the highest yield in the figure (left panels). }
\label{atifig3}
\end{figure}
%%%%%%%%%%%%%%%%%%%%%%%
The ATI angular distributions, depicted in Fig.~\ref{atifig3},
confirm this trend. In fact, the figure shows distributions
symmetric upon $\theta \rightarrow \pi-\theta$ and $\phi \rightarrow
\pi-\phi$, in which it is possible to identify three distinct energy
regions. Near the threshold ($N=15$ and $N=32$ for the lower and
higher intensity, respectively), there exist annular maxima,
centered around $\phi=\pi/2$. Such maxima are most pronounced for
$\theta=0$ and $\theta=\pi$, but also occur for other alignment
angles. As the photoelectron energy increases (middle and right
panels in the figure), the ring-shaped maxima split and move towards
$\phi=n\pi$ and $\theta=n\pi$, which are the maxima observed for
high enough ATI orders. Other, less pronounced maxima are also seen,
as well as additional substructure. Such a pattern is due to
interference effects and comes from the non-trivial part of the
action. As the driving-field intensity increases, this pattern gets
more complicated, but the overall behavior remains.

 \subsection{High-harmonic generation}
Subsequently, we address the question of whether the effects
reported in \cite{atthhg2004,dpg}, for HHG in a near-infrared laser
field superposed to a time-delayed attosecond-pulse train, are
present in our framework. Specifically, intensity variations from
one to two orders of magnitude in the plateau harmonics have been
observed, whose prominence, and even existence, was strongly
affected by the time delay $\phi=\omega t_d$. Moreover, the
resolution of the harmonic peaks, and the plateau structure, with
enhancements of either \emph{all} plateau harmonics, or only near
its low-energy end, were observed to depend very much on this
parameter.

In \cite{atthhg2004}, such effects have been attributed to
quantum-trajectory selection: In HHG, in the absence of the
attosecond pulses, there exists one set of trajectories, which
mainly contribute to the yield. Such trajectories merge at the
cutoff, i.e., the maximal energy for which HHG occurs, and
correspond to a longer and a shorter excursion time for the electron
in the continuum \cite{saclay96}. By using an attosecond-pulse
train, one is splitting this set into two, due to the fact that the
electron is released in the continuum with non-vanishing velocity,
and, at the same time, selecting a trajectory of this set, by fixing
the electron emission time.

Thus, the attosecond pulses allow one to perform quantum-trajectory
selection at the single-atom response level. For solely an IR laser
field, it is only possible to perform such a selection at the
macroscopic level, by controlling the propagation conditions
\cite{saclay96}. In support of this argument, classical simulations
have been performed in \cite{atthhg2004} for an electron returning
to its parent ion. Thereby, the initial electron velocity has been
estimated from the average frequency in the attosecond-pulse train,
which in \cite{atthhg2004}, has been composed by five harmonics
($N=11$ to $N=19$).

\begin{figure}[tbp]
\begin{center}
\hspace*{-1cm}
\includegraphics[width=9.5cm]{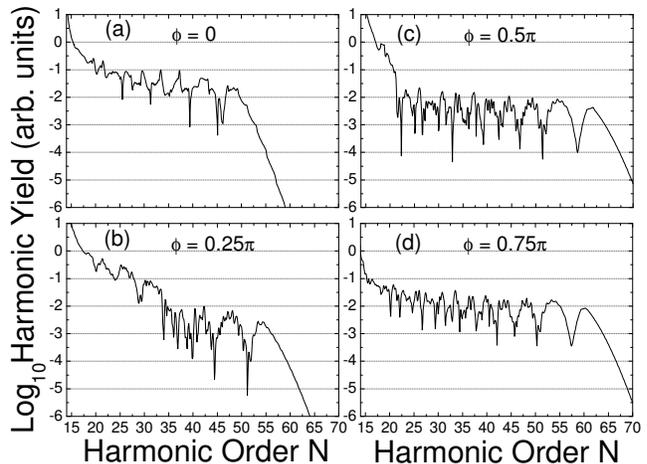}
\end{center}
\caption{High-harmonic spectra for neon ($I_p=0.79$ a.u.) interacting with
a monochromatic field of intensity $I=5\times 10^{14}
\mathrm{W/cm^2}$, and frequency $\omega=0.057$ a.u., and a train of
attosecond pulses of intensity
$I_{\mathrm{h}}=10^{13}\mathrm{W/cm^2}$ (Eq. (\ref{attotrain})), and
several time delays between the attosecond pulse train and the IR
field. We considered the orbits with the five shortest pairs of
excursion times $t-t^{\prime}$. In the figure, we only show the
harmonic orders for which $\Omega>I_p$.} \label{hhgfig1}
\end{figure}

In Fig.~\ref{hhgfig1}, we present HHG spectra computed with Eq.
(\ref{hhgatto}), in the broad gaussian approximation. Such spectra
are highly dependent on the time delay $t_d$ between the
low-frequency field and the attosecond-pulse train, and differ in
the cutoff energy, and in the high-harmonic strengths and
resolutions. Specifically, in the examples provided in the figure,
the cutoff energies vary from roughly $I_p+1.8U_p$ [Fig.
\ref{hhgfig1}.(a)] to $I_p+2.5U_p$ [Fig.~\ref{hhgfig1}.(c)], and the
intensities of the plateau harmonics differ in up to two orders of
magnitude. Furthermore, for particular time delays, the plateau
harmonics are unequally enhanced. For instance, in
Figs.~\ref{hhgfig1}.(b) and
 \ref{hhgfig1}.(c), the low-plateau harmonics are more than one order
of magnitude larger than those near its high-energy end. Such a
group varies from the lower half [Fig.~\ref{hhgfig1}.(b)] to the
lowest few harmonics [Fig.~\ref{hhgfig1}.(c)] of the plateau. The
spectra are identical over $\omega t_d \rightarrow \omega t_d \pm
\pi$ (not shown). This is a consequence of the fact that $\left|
d_{\Omega}(\omega t_{d})\right| ^{2}=\left| d_{\Omega}(\omega
t_{d}\pm \pi )\right| ^{2}$. In particular, for $\omega t_d=n\pi$,
the plateau harmonics exhibit maximal intensities in this context.

%%%%%%%%%%%%%%%%%%%%%%%%%%%%%%%
%%%%%%%%%%%%%%%%%%%%%%%%%%%%%%%%%%%%%%%%%%%%%%%5

The effects observed in Fig.~\ref{hhgfig1} are in qualitative
agreement with those reported in \cite{atthhg2004,dpg}. In order to
find out whether they are due to quantum-orbit selection, they will
be analyzed in terms of the orbits of an electron recombining with
its parent ion. For that purpose, we solve Eq. (17) for a fixed
start time. The real parts of its solutions can be directly
associated to the return times $t$ of a classical electron in an
external field.

\begin{figure}[tbp]
\hspace*{-1cm}
\includegraphics[width=9.5cm]{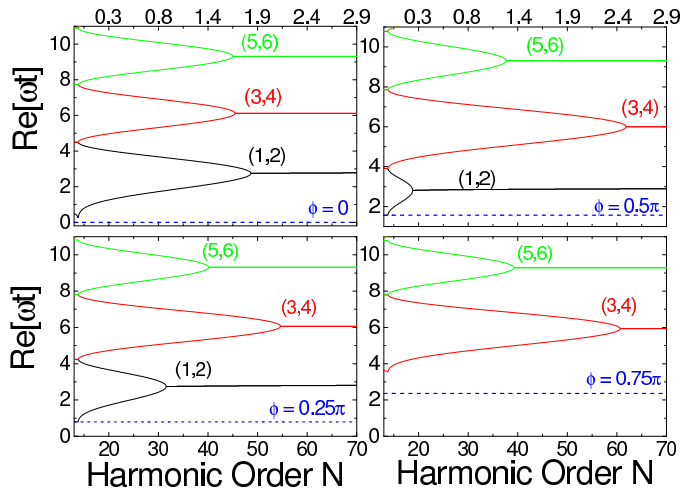}
\caption{(Color online) Real part of the recombination
times as a function of the high-harmonic order, for neon ($I_p=0.79$
a.u.) interacting with a monochromatic field of intensity $I=5\times
10^{14} \mathrm{W/cm^2}$, and frequency $\omega=0.057$ a.u. and an
attosecond-pulse train of intensity
$I_{\mathrm{h}}=10^{13}\mathrm{W/cm^2}$, composed of an infinite
number of harmonics (Eq. (\ref{attotrain})). Parts (a), (b), (c) and
(d) correspond to delays $\phi=0$, $\phi=0.25\pi$, $\phi=0.5\pi$ and
$\phi=0.75\pi$ between the attosecond-pulse train and the
low-frequency driving wave, respectively. In the figure, we only
show the harmonic orders for which $\Omega>I_p$, and the pairs of
orbits are indicated by the natural numbers $(i,j)$, which increase
according to the electron excursion time in the continuum. The ionization
times are indicated by the blue dashed lines, and the numbers in the
upper part of the figure give the approximate kinetic energy of the
electron upon return, in units of the ponderomotive energy $U_p$.}
\label{hhgfig2}
\end{figure}

Fig.~\ref{hhgfig2} depicts the real parts of such return times for
the six shortest orbits, as functions of the harmonic order, for the
time delays in the previous figure. The orbits are denoted by the
numbers $(i,j)$, which increase with the electron excursion time in
the continuum. Depending on the time delay, the return times, and
the maximal harmonic energies are significantly altered. This shows
that the changes in the initial conditions of the electron being
ejected considerably affect its subsequent dynamics. Furthermore,
the return times occur in pairs, i.e., for a single ionization time,
there is always a shorter and a longer travel time for the electron
in the continuum. Such pairs of orbits coalesce at the maximally
allowed harmonic energies, which considerably vary with the delay
$\phi$. Such variations are explicitly given in Table I. In
particular, the largest variations occur for the set of orbits
(1,2), namely from the ionization potential to $I_p+1.8U_p$.
%%%%%%%%%%%%%%%%%%%%%%%%%%%%%%%
%%%%%%%%%%%%%%%%%%%%%%%%%%%%%%%%%%%%%%%%%%%%%%%5

\begin{table}
\begin{tabular}{ll}
\hline\hline Orbits \hspace*{0.5cm}& $E_{\mathrm{kin}}(t,t^{\prime })/U_{p}$ \\
\hline\hline &
\begin{tabular}{llll}
$\phi =0$ \hspace*{0.7cm}& $\phi =0.25\pi $ \hspace*{0.3cm}& $\phi =0..5\pi $ \hspace*{0.4cm}& $\phi =0.75\pi $%
\end{tabular}
\\ \hline
\begin{tabular}{l}
(1,2) \\
(3,4) \\
(5,6)
\end{tabular}
&
\begin{tabular}{llll}
1.80\hspace*{1cm} & 0.93\hspace*{1cm} & 0.26 \hspace*{1cm}& -- \\
1.65 \hspace*{1cm}& 2.11 \hspace*{1cm}& 2.50 \hspace*{1cm}& 2.43\hspace*{1cm} \\
1.62 & 1.37 & 1.25 & 1.32
\end{tabular}
\\ \hline\hline
\end{tabular}
\caption{Maximum kinetic energies
$E_{\mathrm{kin}}(t,t^{\prime})=(N\omega-I_p)/U_p$, in units of the
ponderomotive energy, for an electron along the three pairs of
orbits displayed in Fig.~\ref{hhgfig2}. No entry means that this
energy is very close to the ionization potential.}
\end{table}

 Each of such local maxima
potentially corresponds to a cutoff in the harmonic spectra.
Whether,
 however, they will lead to only minor features or to a sharp decrease
in the high-harmonic yield will depend on three main issues. First,
it is necessary that the electron reaches the continuum with a large
ionization probability. Second, due to the fact that the electronic
wave packet spreads with time, it is desirable that such a wave
packet spends as little time as possible in the continuum. This
yields a large overlap between such a wave packet and that of a
bound electron, and, consequently, prominent harmonics. Finally,
there may be effects related to quantum interference, which suppress
or enhance particular sets of harmonics. Since, due to the
attosecond-pulse train, the electron is being ejected with a large
probability for all cases, we expect that the main difference will
be due to wave-packet spreading. Hence, the shorter the electron
excursion time $\tau=t-t^{\prime}$ in the continuum is, the more
prominent the corresponding harmonics should be.

\begin{figure}[tbp]
\hspace*{-1cm}
\includegraphics[width=10cm]{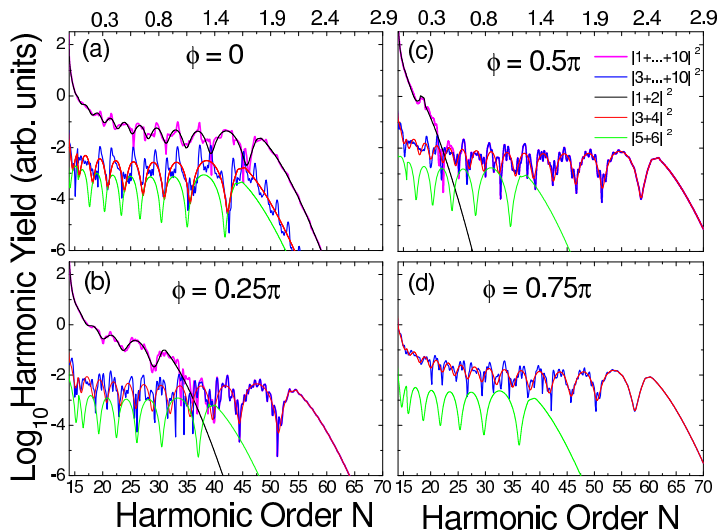}
\caption{ (Color online) Contributions of isolated pairs of orbits
to the high-harmonic spectra for the same atomic and external-field
parameters in Figs.~4 and 5, and delays $\phi=0$, $\phi=0.25\pi$,
$\phi=0.5\pi$ and $\phi=0.75\pi$ between the attosecond-pulse train
and the low-frequency driving wave
(Figs.~\ref{hhgfig3}(a),~\ref{hhgfig3}.(b),~\ref{hhgfig3}.(c)
and~\ref{hhgfig3}.(d), respectively). In the figure, we only show
the harmonic orders for which $\Omega>I_p$. The contributions from
specific pairs of orbits are depicted in the same colors as those
employed in Fig. 5, for depicting each corresponding pair. For
comparison, the spectra computed with the ten shortest orbits and
with orbit 3 to 10 are given as the thick lines in the figure.  The
numbers on the upper parts of the figure give the approximate
kinetic energy of the electron upon return, in units of the
ponderomotive energy.} \label{hhgfig3}
\end{figure}

We will now analyze the contributions of each pair of orbits to the
spectrum. Such contributions are presented in Fig.~\ref{hhgfig3}.
Panel \ref{hhgfig3}.(a) shows that, for an attosecond-pulse train in
phase with the low-frequency field, the main contributions come from
the orbits $(1,2)$. This occurs for two main reasons. First, the
harmonics from this set of orbits are two orders of magnitude larger
than those from the remaining pairs. This is a consequence of the fact
that
 $\tau$ is very short for such a pair.
Furthermore, for this specific phase, the cutoff energy for $(1,2)$
extends to a relatively large harmonic order (c.f. Fig.
\ref{hhgfig2} and Table 1), so that the contributions from the
remaining pairs of orbits are overshadowed.

For a delay of $\phi=0.25\pi$ [Fig.~\ref{hhgfig3}.(b)], this picture
starts to change. In fact, although the harmonics from (1,2) are
much stronger than those from the remaining pairs, the maximal
kinetic energy for $(1,2)$ lies near a much lower harmonic order
(near $N=35$). Thus, beyond such energy, the corresponding
contributions are exponentially decaying and the harmonics
intensities, as well as the cutoff energy near $2.11U_p$, are mainly
determined by $(3,4)$. This causes the double plateau structure in
the figure, and provides an alternative explanation for the
enhancements only in the low-plateau harmonics reported in
\cite{atthhg2004}.

For $\phi=0.5\pi$ [Fig.~\ref{hhgfig3}.(c)], instead, there is a very
steep intensity drop at the the low-energy end of the plateau. This
is a consequence of the very low cutoff for the orbits $(1,2)$, near
the $19^{\mathrm{th}}$ harmonic. For higher energies, the main
behavior in the spectrum is determined by $(3,4)$. Finally, for
$\phi=0.75\pi$ [Fig.~\ref{hhgfig3}.(d)], the maximal energy for
$(1,2)$ is so close to the ionization potential, that such a pair
does not play any role. Consequently, there is an overall decrease
of at least one order of magnitude in the spectrum. At the same
time, the orbits (3,4) become increasingly shorter. Around $\omega
t_d=0.75 \pi$, such latter orbits have become the shortest
contributing and the intensity starts to increase, until, at $\omega
t_d=\pi$, the same pattern as for vanishing time delay is recovered.
The contributions from pairs of orbits with longer excursion times
are much less relevant due to wave-packet spreading.

 Structured
spectra, however, are only obtained if several pairs of orbits are
taken into account. One should note that, if one employs orbits
starting from (3,4) and vanishing time delays between the
attosecond-pulse train and the IR field, there seems to be an
improvement in the resolution of the spectra (c.f. blue line in
Fig.~\ref{hhgfig3}.(a)). In \cite{atthhg2004}, resolved harmonics
have been related to selecting a particular trajectory for the
returning electron. In our case, however, it appears to be an
interference effect between several orbits for the returning
electron. The fast oscillations observed here are possibly not
harmonics (due to the periodicity of the process), but are caused by
the interference of different quantum orbits originating from
(launched from) the same half cycle.

With regard to the above discussion, we should point out that the
absence of discrete harmonics in the spectra displayed in this work
is due to the fact that we are considering orbits starting only in
the first cycle of the laser field, instead of over many cycles.
Therefore, we do not have enough periodicity to obtain discrete
harmonic peaks. This is however not a problem in our context, since
we concentrate our investigations on how quantum-interference
effects influence the main structure of the spectra. By considering
a restricted range of electron start times, such effects can be more
clearly seen and interpreted.

 %%%%%%%%%%%%%%%%%%%%%%%%%%%%%%%
%%%%%%%%%%%%%%%%%%%%%%%%%%%%%%%%%%%%%%%%%%%%%%%

The above-discussed behavior is very similar to that observed in
\cite{atthhg2004}. Figs.~\ref{hhgfig1}-\ref{hhgfig3}, however, have
been plotted for a set of time delays different from those employed
in such a reference. Such discrepancies may be related to the fact
that, in \cite{atthhg2004}, only a finite and relatively small group
of harmonics has been used, whereas, in our computations, we have
employed the attosecond-pulse limit for an infinite set of
harmonics.

In fact, if the attosecond-pulse train is composed by a relatively
small number of harmonics (such as in
\cite{atthhg2004}), it is possible to define a main frequency for
this train. Hence, it is reasonable to assume that electron is being
ejected in the continuum by a photon of such a frequency, and thus
with nonvanishing velocity. This will cause a splitting of the
trajectories into the ``downhill" and ``uphill" transitions, which
correspond to the ejection of the electron in- or opposite to- the
direction of the laser field, respectively, and which will be
selected according to the time delay.

On the other hand, if the attosecond pulse train is composed by an
infinite number of frequencies, the uncertainty relation will no
longer allow one to define a main frequency for it. Consequently,
the physical picture of an electron being released by a photon of a
specific frequency and reaching the continuum with a constant
velocity is no longer applicable. However, it is still possible to
define a specific time $t^{\prime}$ for which the electron is
ejected, and determine the corresponding return times.
\section{Conclusions}
\label{concl}

 In conclusion, we have shown that an attosecond-pulse
train superposed to a strong, near-infrared laser field can be used
for controlling above-threshold ionization (ATI) and high-order
harmonic generation (HHG). This control is exerted by using the
attosecond pulses, which are time delayed with respect to the
infrared field, to constrain the instant in which an electronic wave
packet is ejected in the continuum. Since this wave packet will
either recombine with its parent ion, generating high harmonics, or
contribute to the above-threshold ionization spectrum, both
phenomena are strongly affected by such pulses.

Similar investigations, which led to similar effects to those in
this paper, have been performed in \cite{atthhg2004,attati2005},
employing the numerical solution of the Schr\"odinger equation. In
this work, however, we go beyond such studies, and trace all
features observed to quantum-interference effects between the
possible pathways of the electronic wave packet. In particular, the
model used in this paper, based on the Strong-Field Approximation
(SFA) allows one to investigate the different ingredients involved
in the computation of the transition amplitude, and how they are
influenced by the attosecond-pulse train, in a detailed and
transparent way.

 For ATI,
we restrict ourselves to the direct electrons, i.e., those that
reach the detector without rescattering with its parent ion. We
consider an attosecond-pulse train composed by an infinite group of
harmonics and we sum over all periods of the driving field.
 In this case, we observed an extension in the photoelectron energy if the
attosecond-pulse train is in phase with the IR field (i.e., for
delays $\phi=n\pi$).

In \cite{attati2005}, this has been attributed to the fact that, in
this case, the electrons are ejected at a zero crossing of the IR
field. Hence, they gain the maximal possible momentum from the
field, namely $p=\pm 2 \sqrt{U_p}$. In this paper, we have shown
that this mechanism does play a role, which, in our framework, is
implicit in the non-trivial part of the action. However, in order to
obtain the above-mentioned effect, the dipole matrix elements from a
1s hydrogenic state are also a necessary ingredient.
Physically, such matrix elements are related to the initial state
where the electron is bound and, in the SFA, contain all the
information about the atomic bound potential. Furthermore, we have
shown that such an effect only occurs if the photoelectron energy is
sufficiently far from the ionization threshold. In fact, for low ATI
orders, maximal intensities are obtained for $\phi=(2n+1)\pi/2$.
This is also due to an interplay between the phase the electron
acquires when it reaches the continuum and its initial bound state.

Apart from such studies, we have observed that the ATI angular
distributions are strongly dependent on this delay, and on the
photoelectron energy. Near the ionization threshold, such
distributions exhibit very pronounced maxima for backscattered
electrons (alignment angle $\theta=n\pi$) which leave at peak-field
times ($\phi=(2n+1)\pi/2$). As the photoelectron energies increase,
such maxima move towards the zero crossings of the electric field
($\phi=2n\pi$). This behavior is in agreement with that observed for
the ATI spectra.

 For HHG, depending on the time delay between the
 attosecond-pulse train and the IR field, we observed deviations in
 almost $30\%$ in the cutoff energy.
 Furthermore, the plateau structure is highly sensitive to the time delay $\phi$.
 Indeed, we found that, for certain ranges of this parameter, the
 low-plateau harmonics are one to two orders of magnitude stronger
 than those near its high-energy end. The precise sets of
 harmonics in such a double-plateau structure depend very much on
 $\phi$. For our particular parameters, the set of strong
 harmonics moved towards lower energies as $\phi$ distanced itself
 from a zero crossing of the electric field.

 Our results can be traced back to the
 interplay of two sets of orbits for an electron returning to its
 parent ion.  Specifically, the maximal return electron energy in the shortest set of orbits
  changes from 0 to $1.8U_p$, as the ejection time
  of the electron is varied. These orbits yield a group of very
  strong harmonics, whose extension can be manipulated
  by controlling the time delay $\phi$. The remaining harmonics are
  mainly determined by the second shortest pair of orbits, and are
  considerably weaker.

 In this context, a noteworthy aspect is
 that, in contrast to ATI, for which the effects mentioned in this paper
 also depend on the initial electron state, in HHG they
 appear to be due to controlling electron orbits with
 an attosecond-pulse train. Indeed, whereas in ATI such effects
 do not occur for initial localized bound states, i.e., in the broad
 gaussian approximation, for HHG they are already present in this case.

The results for HHG also agree to a great extent with those reported
in \cite{atthhg2004}. However, their interpretation, as well as the
time delays for which they occur, are somewhat different. In fact,
in \cite{atthhg2004}, all features are explained by selecting a
single orbit from the shortest pair, and by the splitting in this
set of orbits due to the fact that the electron is reaching the
continuum with non-vanishing velocity. Such differences are possibly
due to the fact that, in \cite{atthhg2004}, only a small group of
harmonics has been employed for producing the attosecond-pulse
train, while, in Eq. (\ref{attotrain}), we consider the limit of an
infinite number of harmonics. We therefore expect the agreement to
increase if a larger number of harmonics is taken when constructing
the attosecond-pulse-train.

\section{Appendix}

In this appendix, we explicitly provide the time-dependent
expression adopted in Ref. \cite{hhgsfa}. In this case, the
time-dependent dipole moment reads
\begin{eqnarray}
d(t)&=&i\int_{-\infty }^{t}dt^{\prime }\int d^{3}kd_z^{\ast
}(\mathbf{k}+\mathbf{A}(t))d_z(\mathbf{k}+\mathbf{A}(t^{\prime
}))\nonumber \\&& E(t^{\prime })\exp [iS(t,t^{\prime },
\mathbf{k})], \label{timelhhg}
\end{eqnarray}
with the action
\begin{equation}
S(t,t^{\prime },\mathbf{k})=-\frac{1}{2}\int_{t^{\prime }}^{t}[%
\mathbf{k}+\mathbf{A}(\tau )]^{2}d\tau -I_{p}(t-t^{\prime }).
\label{actionhhg1}
\end{equation}

Alternatively to the procedure adopted in this paper, one may
integrate Eq.(\ref{timelhhg}) over the intermediate momentum using
the saddle point equation (\ref{saddle3}). This yields the equation
\begin{eqnarray}
b_{t}&=&\frac{i\pi E_{h}}{\sigma}\sum_{n=0}^{\infty}
(-1)^{n}\int_{-\infty }^{t}\hspace*{-0.5cm}(2\pi/i(t-t^{\prime}_n+ i
\epsilon))^{3/2}\\&& d_z^{\ast
}(\mathbf{k}_{st}+\mathbf{A}(t))d_z(\mathbf{k}_{st}+\mathbf{A}(t^{\prime
}_n))\nonumber \exp [iS(t,t^{\prime }_n ,\mathbf{k}_{st})],
\label{timehhg}
\end{eqnarray}
with the action given by formula (20), to which a Fast Fourier
transform routine is then applied. In Eq. (21), $\mathbf{k}_{st}=-
(1/(t-t^{\prime }_n))\int_{t^{\prime }_n}^{t}d\tau \mathbf{A}(\tau
)$, and the sum extends over $t^{\prime }_n \le t$. Due to the
wave-packet-spreading effects (described by the
($\left(2\pi/i(t-t^{\prime}_n+ i \epsilon) \right)^{3/2}$), the main
contributions to the sum in Eq. (21) come from few $t^{\prime }_n$
that are the closest to $t$.

\acknowledgements This work has been financed in part by the
Deutsche Forschungsgemeinschaft (European Graduate College
``Interference and Quantum Applications", SFB407, SPP1078 and
SPP1116, 436POL)), spanish MEC grant FIS-2005-04627, the ESF
Program QUDEDIS, and the Marie Curie European Program MRTN-CT-2003-505138,XTRA.
C.F.M.F. would like to thank the ICFO and the
Universidade do Algarve for their kind hospitality.


\begin{thebibliography}{99}
\bibitem{footnsfa} The SFA consists in neglecting
the atomic potential in the propagation of the electron in the
continuum, the laser field when the electron is bound or rescatters,
and the excited states of the atom in question.
\bibitem{footnfinite}For the parameter range employed in this paper,
we obtained practically identical results from such a limit, and
from Eq. (\ref{atifinite}), for $\zeta \lesssim 0.04$.
\bibitem{footnote} For this latter intensity, tunneling ionization
may also play a role and compete with the ionization induced by the
attosecond pulse train. Hence, it may be necessary to take into
account tunneling ionization in the transition amplitude
(\ref{atiampl}). This case is, however, considered to illustrate the
effects that may arise at high laser intensity.
\end{thebibliography}
\end{document}